# The Physics of Unbounded, Broadband Absorption/Gain Efficiency in Plasmonic Nanoparticles

Nasim Mohammadi Estakhri, and Andrea Alù

*Department of Electrical and Computer Engineering, The University of Texas at Austin, Austin, TX 78712, USA*

Anomalous resonances in properly shaped plasmonic nanostructures can in principle lead to infinite absorption/gain efficiencies over broad bandwidths. By developing a closed-form analytical solution for the fields scattered by conjoined semicircles, we outline the fundamental physics behind these phenomena, associated with broadband adiabatic focusing of surface plasmons at the nanoscale. We are able to justify the apparent paradox of finite absorption/gain in the limit of infinitesimally small material loss/gain, and we explore the potential of these phenomena in nonlinear optics, spasing, energy-harvesting and sensing.

The growing interest in the optical properties of nanoparticles [1]-[2] has led to the discovery of many counterintuitive scattering features in plasmonic nanostructures. Due to their negative real part of permittivity, these particles support surface plasmon resonances at the nanoscale that have been proposed for many exciting applications, including field concentration, sensing, nanolasing and optical guiding [3]-[8]. Different configurations have been analyzed in recent years, from simple nanospheres and core-shell structures [6]-[7] to more complicated shapes, like crescent-shaped cylinders [9]. If simple configurations are known to support strong, sharp plasmon resonances, more complicated shapes may provide more complex scattering responses, such as Fano and EIT resonances [10]-[11], or broadband operation [9]. Including gain may further boost these effects and compensate the detrimental effects usually caused by losses [12]-[13]. Many of the exotic properties of these geometries, however, often appear to contradict well-established physical limitations of resonant subwavelength systems [14], and the underlying physics is often difficulty captured because of the complex interaction between multiple resonances and plasmonic effects.

As an example that may shed significant new light into these phenomena, we analyze here the anomalous electromagnetic response of a rather simple composite nanoparticle, formed by two conjoined half-cylinders of arbitrary complex permittivity $\varepsilon_1$, $\varepsilon_2$ relative to the background permittivity, and radius $a$, as shown in the inset of Fig. 1. This geometry has been recently proposed in the special configuration $\varepsilon_1 = -\varepsilon_2$ to form a resonant optical nanocircuit and previous attempts to analytically solve its scattering properties using mode-matching analysis [15], integral transformations [16] and coordinate mapping [17] have led to nonphysical solutions and strong numerical instabilities. We show in the following that these challenges are associated with remarkably counterintuitive resonant phenomena, which lead to a continuous frequency range over which distributed plasmon resonances may support unbounded values of absorption or gain efficiency, i.e., finite absorption or gain even in the limit of infinitesimally small material loss/gain. By extending the analytical approach originally introduced in [17] to evaluate the polarizability of a hemicylinder, we are able to solve in closed-form the complete scattering problem associated with this geometry and derive closed-form expressions for the induced fields inside and outside this composite particle. This solution provides valuable physical insights into the complex wave interaction of this particle over a broad range of frequencies, which may provide, as we discuss in the following, exciting possibilities for energy concentration, harvesting and spasers [18]-[19]. We also relate these findings to other related plasmonic geometries that may lead to a similar anomalous response.

We start by solving the scattering problem in the quasi-static limit, under the assumption $a << \lambda_0$. An incident monochromatic wave with electric field $\mathbf{E}_0$ illuminates the nanostructure under an $e^{j\omega t}$ time convention and the permittivities of the two half-cylinders can take arbitrary complex values, whose imaginary parts correspond to material loss or gain depending on their negative or positive sign. Due to symmetries and linearity, the problem may be split into two orthogonal excitations with respect to the common diameter of the structure. By using separation of variables in the 2D bipolar coordinate system [17], the potential distribution in each material may be written in integral form as

$$\varphi_i(u,v) = \int_0^\infty U(u)\left[C_{i1}(\lambda)\cosh(\lambda v) + C_{i2}(\lambda)\sinh(\lambda v)\right]d\lambda \tag{1}$$

in which the subscript $i = 1,2,0$ refers to upper, lower and outer regions, respectively, $\lambda$ is the continuous eigenvalue, $U(u)$ is either $\cos(\lambda u)$ or $\sin(\lambda u)$ for longitudinal and transverse polarizations respectively, and $-\infty < u < \infty$, $-\pi < v \leq \pi$ are bipolar coordinate variables. The unknown coefficients $C_{ij}(\lambda)$ may be found by applying suitable boundary conditions at the various boundaries to calculate the general form of potential distribution in all space from Eq. (1). In [17], this integral expansion was used to determine the electric polarizability $\alpha = p/E_0$ of an isolated hemicylinder, where $p$ is the induced electric dipole moment, evaluated using the asymptotic expression of $\varphi_0$ in the far-field.

In the present case of two conjoined hemicylinders, the normalized polarizability may be analogously derived for arbitrary relative permittivities. For longitudinal excitation its value is

$$\begin{aligned}\alpha_l &= \frac{\pi^2\left[-\varepsilon_2 + \varepsilon_1(-1 + 6\varepsilon_2)\right] + 12(\varepsilon_1 + \varepsilon_2)\left(\mathrm{Li}_2(\varepsilon^-) + \mathrm{Li}_2(\varepsilon^+)\right)}{1.5\pi^2(\varepsilon_1 + \varepsilon_2 + 2\varepsilon_1\varepsilon_2)} \\ \varepsilon^\pm &= -\frac{(1+\varepsilon_1)(1+\varepsilon_2)(\varepsilon_1 + \varepsilon_2)}{\varepsilon_2 \pm \sqrt{-(\varepsilon_1-\varepsilon_2)^2(2+\varepsilon_1+\varepsilon_2)(\varepsilon_1+\varepsilon_2+2\varepsilon_1\varepsilon_2) + \varepsilon_1\left[1+\varepsilon_2(4+\varepsilon_1+\varepsilon_2)\right]}}\end{aligned} \tag{2}$$

in which $\mathrm{Li}_2(x)$ is the polylogarithm function of second order, and a similar expression can be derived for the transverse case [20]. Fig. 1 shows the calculated polarizabilities in the case $\varepsilon_2 = 3$ for different values of $\varepsilon_1$, assuming lossless materials (real-valued $\varepsilon$). Since in the quasi-static limit radiation losses are neglected, in the limit of no Ohmic absorption the extracted power $P_{ext} = -\omega/2|E_0|^2 \mathrm{Im}[\alpha]$ should be identically zero, which requires the polarizability to be purely real. On the contrary, the results in Fig. 1 highlight two continuous frequency ranges (yellow shades in the figure) over which both polarizabilities have an imaginary component even in this lossless limit, consistent with some of the findings in [17] for a single hemicylinder. This paradox is mathematically associated with the range of permittivities for which the arguments of $Li_2$ have magnitude larger than one, which requires to analytically continue the polylogarithm in the complex plane. Two branch-cuts are associated with this range of complex solutions, and complex conjugate values are admissible solutions of (2). This implies that our geometry may be able to extract (or produce, depending on the sign of $Im[\alpha]$) power, even in the case of purely lossless (or gainless) materials. In Fig. 1 we indicate with solid (dashed) lines the solution with $Im[\alpha]<0$ ($Im[\alpha]>0$), corresponding to absorbing (lasing) nanoparticles [21].

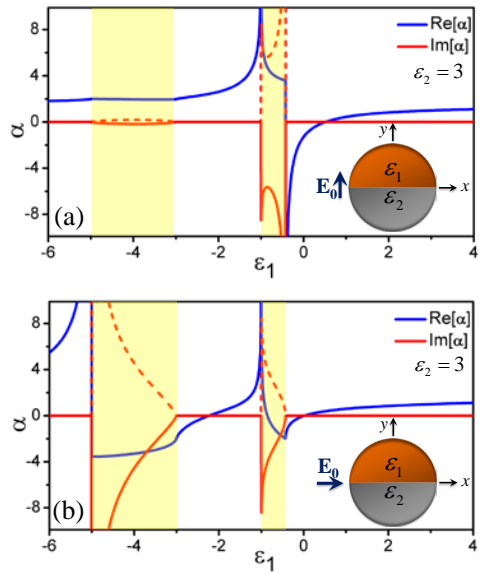

FIG. 1 Complex polarizability of two conjoined half-cylinders with $\varepsilon_2 = 3$ (a) for longitudinal and (b) transverse excitation. Particle geometry and excitation fields are shown in the inset.

As shown in the following, this counterintuitive response is physically associated with the singularities induced at the two corners of the structure, which have so far been assumed as ideal mathematical edges with zero curvature at the tip. Not surprisingly, the permittivity range over which $Im[\alpha] \neq 0$ exactly corresponds to the quasi-static eigen-resonance of a 90° double dielectric wedge, which arises for $\varepsilon_2 > 0$ when [20],[22]:

$$-\varepsilon_2 - 2 < \varepsilon_1 < \min\{-\varepsilon_2, -1\}, \quad \max\{-\varepsilon_2, -1\} < \varepsilon_1 < \frac{-\varepsilon_2}{1 + 2\varepsilon_2}. \tag{3}$$

Similar inequalities may be easily derived for $\varepsilon_2 < 0$ [20], providing in general two separate continuous resonant windows of unbounded absorption/gain efficiency, defined as the ratio $\text{Im}[\alpha]/\varepsilon_i$ with $\varepsilon_i$ being $\text{Im}[\varepsilon_1]$ or $\text{Im}[\varepsilon_2]$. In the permittivity range (3), the corners support continuous eigenmodes that are at the basis of this anomalous response. In the special case of a hemicylinder $\varepsilon_2 = 1$ [15], the two windows merge into $-3 < \varepsilon_1 < -1/3$, separated by a single point $\varepsilon_1 = -1$, corresponding to the special internal resonance analyzed in [23].

In the ideal lossless limit, there is no way to distinguish between the two branch-cuts, and both conjugate solutions in Fig. 1 are equally admissible. This implies that the boundary-value problem is not well defined, as the uniqueness theorem does not apply to this ideal lossless scenario [24]. Small losses are required to select the correct Riemann sheet and assign proper meaning to the solutions in Fig. 1. In order to address this issue, Fig. 2 shows the effect of loss/gain in $\varepsilon_1$ on $\text{Im}[\alpha]$ for different values of $\text{Re}[\varepsilon_1]$. Outside the resonance region, e.g., $\text{Re}[\varepsilon_1] = -6.5$ (blue lines in Fig. 2), $\text{Im}[\alpha]$ is a well-behaved continuous odd function of $\text{Im}[\varepsilon_1]$, and is identically zero for zero material loss. For values that lie in the continuous resonant range (3), $\text{Im}[\alpha]$ is still an odd function of $\text{Im}[\varepsilon_1]$, but it has a discontinuity at $\text{Im}[\varepsilon_1] \to 0^{\pm}$, associated with the ambiguity in selecting the correct Riemann sheet in the lossless case. By introducing an arbitrary amount of loss $\varepsilon_i < 0$ or gain $\varepsilon_i > 0$, we are able to select either the absorptive ($\text{Im}[\alpha] < 0$) or lasing ($\text{Im}[\alpha] > 0$) branch in Fig. 2. Interestingly, this implies that an arbitrarily small (but mathematically nonzero) value of loss or gain in the material can provide finite absorption or lasing over a continuous bandwidth corresponding to (3). In this continuous range, absorption or gain efficiencies are effectively unbounded. Interestingly, smaller absorption/gain in the material can lead to larger overall absorption/gain in the nanoparticle, as the plasmonic effect at the corner is less quenched.

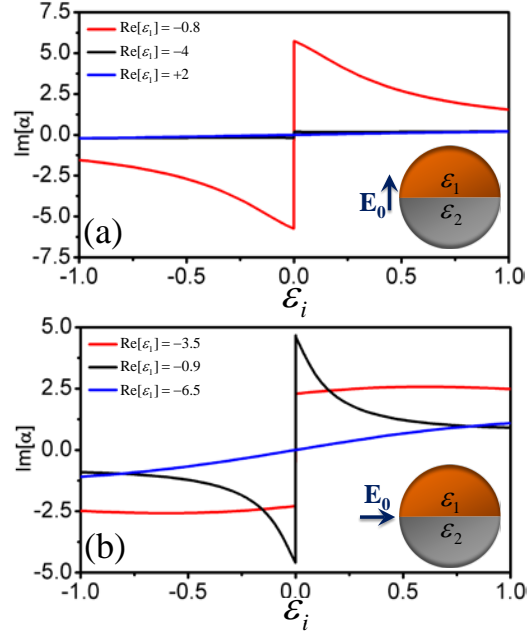

FIG. 2 Imaginary part of polarizability versus $\text{Im}[\varepsilon_1]$ for different values of $\text{Re}[\varepsilon_1]$ for (a) longitudinal and (b) transverse excitation.

Power extraction or generation can arise only in regions where the quadrature component of the potential $\text{Im}[\varphi_i]$ is nonzero. In the quasi-static lossless limit, we would expect $\varphi_i$ to be exactly in phase with the excitation at all points, but in the resonant range (3), analogous to (2), we need a nonzero imaginary component to justify power extraction. In order to gain further insights into this anomalous phenomenon, we have been able to calculate in closed-form also the imaginary part of the potential distributions everywhere (1). As expected, we found $\text{Im}[\varphi_i] = 0$ outside the resonance range in the lossless limit. However, in the resonant range (3) the coefficients $C_{ij}(\lambda)$ have a simple pole in the denominator which causes a Dirac-$\delta$ distribution in the imaginary part, which allows calculating the imaginary part of (1) and (2) in closed form [20]. In the case of a hemi-cylinder $\varepsilon_2 = 1$ and longitudinal excitation, for instance, we find the potential distribution in the upper half-cylinder to be:

$$\mathrm{Im}[\varphi_1] = E_0 \xi \frac{\cos(\lambda_0 u)}{\sinh(\lambda_0 \pi)} \left[ \left( -2\coth(\lambda_0 \pi) + \frac{\xi - 2}{\sinh(\lambda_0 \pi)} \right) \sinh(\lambda_0 (\pi - v)) + (\xi - 1) \frac{\sinh(\lambda_0 (\pi/2 - v))}{\sinh(\lambda_0 \pi/2)} \right], \quad (4)$$

$$\lambda_0 = \frac{1}{\pi} \cosh^{-1}(\xi^2/2 - 1), \xi = (\varepsilon_1 - 1)/(\varepsilon_1 + 1)$$

and similar expressions may be derived for arbitrary values of $\varepsilon_1, \varepsilon_2$ and for the potential distribution in every point in space. Similarly, we can write the imaginary part of longitudinal polarizability (2) in the simple closed form:

$$\mathrm{Im}[\alpha] = \frac{-16}{\pi} \frac{\varepsilon_1 + 1}{3\varepsilon_1 + 1} \cosh^{-1}\left( \frac{(\varepsilon_1 - 1)^2 - 2(\varepsilon_1 + 1)^2}{2(\varepsilon_1 + 1)^2} \right) \mathrm{sign}\left( \frac{\varepsilon_1 - 1}{\varepsilon_1 + 1} \right), \quad (5)$$

which allows writing also in closed form $P_{ext}$. Obviously, Eqs. (4)-(5) are valid only in the resonant range (3) and are zero elsewhere. The sign term in this last equation ensures the proper choice of the branch cut in the lossless limit. By adding an infinitesimally small amount of loss/gain the solution will automatically collapse to the correct branch (Fig. 2).

Figs. 3a-b, as an example, show the real and imaginary parts of the potential distribution for a hemicylinder ($\varepsilon_2 = 1$) with $\varepsilon_1 = -1.1$ and longitudinal excitation. The imaginary part is calculated using (4), whereas the real part is obtained by numerically integrating (1). The imaginary component of the potential represents an eigen-mode of the structure supported by plasmonic resonances at the two corners. This distribution, integrated over the nanoparticle volume, effectively supports the extracted/generated power. It should be noted that in the ideal lossless limit this distribution is not square-integrable, as it leads to a finite value of extracted/generated power for $\varepsilon_i \to 0$. This explains the reason behind the nonuniqueness of the solution in the lossless limit.

Inspecting the imaginary part of the potential distribution in Fig. 3b, we notice strong plasmonic oscillations around the nanoparticle corners. Its variation along the particle diameter is plotted in Fig. 3c, highlighting that the surface plasmon supported by the metal-dielectric interface is adiabatically focused towards the corners, with a finer and finer spatial variation as the corner is approached. This effect, which is supported over the whole resonant range (3), produces broadband, largely enhanced electric fields and it sustains absorption/lasing even for infinitesimally small values of material loss/gain. Essentially, the surface plasmon is adiabatically focused towards the corner, as if it were traveling to infinity, explaining the reason why negligible losses (gain) are sufficient to sustain large absorption (lasing). Different from conventional adiabatic focusing of surface plasmons, in this geometry this effect is achieved at the nanoscale. Field distributions and additional discussions on this effect are reported in [20]. These distributed resonances and adiabatic focusing have direct analogies with the resonant distribution highlighted in [9],[25]-[26], for crescent-shaped and touching plasmonic cylinders, but is obtained here in an arguably simpler geometry over a flat surface.

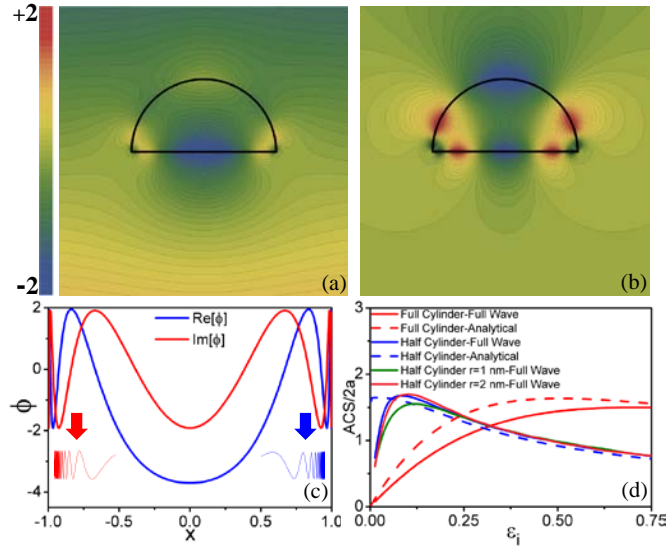

FIG. 3 (a) Real part and (b) imaginary part of the potential distribution for a half-cylinder with $\varepsilon_1 = -1.1$ for longitudinal excitation, normalized to the impinging potential; (c) distribution along the $x$-axis (d) Absorption efficiency vs. material loss for a hemi-cylinder with $\varepsilon_1 = -0.529 - j\varepsilon_i$ compared to full-wave simulations for 1 and 2 nm curvature radii. The full cylinder is also shown for comparison.

The previous analysis highlights that the apparent paradox of unbounded absorption/lasing efficiencies in the proposed nanoparticle is related to two relevant assumptions: ideal singularities in the nanoparticle geometry (perfect corners) and quasi-static solution. In the following we relax both these assumptions and analyze how these effects may be translated into realistic geometries and setups. In the long-wavelength limit, the quasi-static solution can be easily extended to the dynamic regime to include effects of radiation and retardation [27]. The dynamic Mie dipolar coefficient $C_1$ is related to the static polarizability (2) via $C_1 = \left(-1 + j8x_0^{-2}\alpha^{-1}/\pi\right)^{-1}$, $x_0 = k_0 a$, which includes now radiation losses. Fig. 4 shows the absorption efficiency, defined as absorption cross-section normalized to the physical width of the particle, for composite cylinders with $2a = 40 nm$, compared to the case of a homogeneous cylinder of same size. In this case, in order to include also frequency dispersion and realistic material absorption, the upper half-cylinder is chosen to be silver with $\varepsilon_r = \varepsilon_\infty - \omega_p^2 / \omega(\omega - j\Gamma)$, $\varepsilon_\infty = 5$, $\omega_p = 2\pi \times 2175 THz$ and $\Gamma = 2\pi \times 4.35 THz$ [28]. We compare the case of a silver hemi-cylinder ($\varepsilon_2 = 1$) and the case $\varepsilon_2 = 3$, which have different resonant bands following (3). The results confirm that absorption/gain may be largely enhanced over a continuous and controllable frequency band, significantly broadening the range and level of absorption and gain compared to a full circular rod. Similar results are shown in [20] for the case of gold, showing analogous performance. We also further discuss the separate effects of retardation and material loss in [20].

Next, we analyze the effect of finite curvature at the corners. As discussed in [29], when a mathematical edge is replaced by one with nonzero curvature, the continuous eigen-resonance range is necessarily converted into a set of discrete resonance frequencies, which ensures that Chu's fundamental limit is satisfied [14]. The amount of realizable absorption will depend on how adiabatically surface plasmon resonances may be focused and absorbed before the edge is terminated. We used full-wave simulations to study this effect for different curvature values. Absorption efficiency of a blunted hemi-cylinder with permittivity $\varepsilon_1 = -0.529 - j\varepsilon_i$ and $2a = 40 nm$ is compared in Fig. 3d to the ideal geometry. The full cylinder is also shown for comparison. We notice that the absorption phenomenon is pretty robust for finite values of material loss, and the edge bluntness effectively sets the lower level of $|\varepsilon_i|$ for which large absorption/gain may be achieved [20]. In other words, when considering corners with finite curvature, absorption/gain efficiencies are inherently bounded and fundamentally limited by how sharp the corner may be made. Still, significantly large, broadband absorption/gain effects may be achieved with realistic nanogeometries.

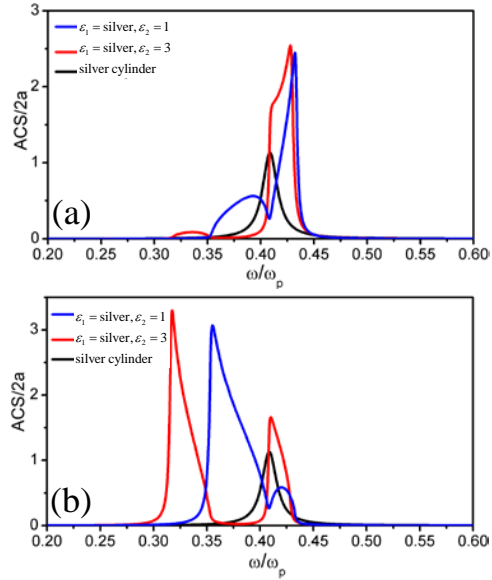

FIG. 4 Absorption efficiency for (a) longitudinal (b) transverse excitation of a composite nanoparticle with upper half-cylinder made of silver and different values of $\varepsilon_2$. The full cylinder case is also shown for comparison.

The distributed resonances and anomalous behavior of the proposed composite nanoparticle may have many exciting applications, including enhanced energy harvesting and spasers [12],[18] based on materials with limited absorption/gain coefficients and an arguably simple configuration from the fabrication point of view. These resonances may be broadband and with a bandwidth and enhancement level controllable by geometry and design. In this sense, we notice that the bandwidth of enhancement is effectively controlled by the corner geometry, and sharper corners can support eigenmodes over even broader continuous bandwidths. We are currently exploring related geometries, such as cylindrical and spherical caps, for which anomalous, broadband enhancement of absorption and gain may be achieved over even larger frequency ranges, based on similar principles. Adiabatic plasmonic focusing at the corners may also be used for other exciting applications, such as sensing and enhanced optical nonlinear effects like switching and nanomemories [11]. The field enhancement may be tailored to be extended all over the particle volume or be confined only around the corners, with exciting implications for these applications [20]. The rapid and sharp variation of absorption versus frequency observed in Fig. 4 can also be used for sensing [30], with sharp linewidths that are comparable to the ones associated with Fano phenomena [10]. Finally, these effects may have a great interest in boosting the usually low values of gain coefficients in natural optical materials, of great interest for loss compensation in metamaterials and plasmonics [13],[31]-[33], as well as for efficient spasers [12],[18]-[19],[34].

This work has been supported by the AFOSR YIP award No. FA9550-11-1-0009 and the ONR MURI grant No. N00014-10-1-0942.

# Supplementary material for the paper "The Physics of Unbounded, Broadband Absorption/Gain Efficiency in Plasmonic Nanoparticles"

In this supplementary material, we expand and provide more details on several concepts presented in our main paper. We first derive the static wedge model of the proposed nanoparticle corners. We then provide the analytical closed-form formulas to calculate the polarizability of a double half-cylinder configuration and comment on the proper choice of branch-cuts in the lossless limit. We also discuss the dynamic behavior of the nanoparticles in more detail investigating the effect of radiation loss versus intrinsic material loss on its resonance absorption. Finally, we show the field and potential distributions at resonance and comment on the effects of blunted (realistic) corners on the overall performance.

## 1. Static Wedge

As discussed in the main text, oscillatory fields adiabatically focusing towards the corners of the composite nanoparticle are responsible for the anomalous absorption/gain occurring in the resonance band. In the corner proximity, the geometry may be modeled as a double dielectric wedge described by Laplace equation. Independent of the polarization of the applied field, eigensolutions may be supported by the wedge configuration for some specific values of material permittivities. It is possible to show that the general resonance condition for a $90°$ double wedge is [1]:

for $\varepsilon_2 > 0$:

$$\left(\max\{-\varepsilon_2,-1\} < \varepsilon_1 < -\frac{\varepsilon_2}{2\varepsilon_2+1}\right) \| \left(-\varepsilon_2-2 < \varepsilon_1 < \min\{-\varepsilon_2,-1\}\right) \tag{S1}$$

for $\frac{-1}{2} < \varepsilon_2 < 0$:

$$\left(-\varepsilon_2 < \varepsilon_1 < -\frac{\varepsilon_2}{2\varepsilon_2 + 1}\right) \| \left(-\varepsilon_2 - 2 < \varepsilon_1 < -1\right) \tag{S2}$$

for $-1 < \varepsilon_2 < \frac{-1}{2}$:

$$\left(-\varepsilon_2 < \varepsilon_1 < \infty\right) \| \left(-\infty < \varepsilon_1 < -\frac{\varepsilon_2}{2\varepsilon_2 + 1}\right) \| \left(-\varepsilon_2 - 2 < \varepsilon_1 < -1\right) \tag{S3}$$

and finally for $\varepsilon_2 < -1$:

$$\left(-\varepsilon_2 - 2 < \varepsilon_1 < -\varepsilon_2\right) \| \left(-1 < \varepsilon_1 < -\frac{\varepsilon_2}{2\varepsilon_2 + 1}\right) \tag{S4}$$

Note that the infinite permittivity range in (S3) does not imply an infinite resonance bandwidth, because of the finite range of permittivity for $\varepsilon_2$ over which it is valid.

## 2. Polarizability in Closed-form

Based on our analytical approach we have the tools to write in closed-form the imaginary part of polarizability in a more compact form compared to Eq. (2) in the main text. Similar to the longitudinal polarization reported in the main text, for hemi-cylinder and transverse polarized excitation we find

$$\mathrm{Im}[\alpha] = \frac{16}{\pi} \frac{\varepsilon_1 + 1}{\varepsilon_1 + 3} \cosh^{-1}\left(\frac{(\varepsilon_1 - 1)^2 - 2(\varepsilon_1 + 1)^2}{2(\varepsilon_1 + 1)^2}\right) \mathrm{sign}\left(\frac{\varepsilon_1 - 1}{\varepsilon_1 + 1}\right) \tag{S5}$$

inside the resonance range using direct integration of the singularity in the integrand. While the term $\mathrm{sign}\left(\frac{\varepsilon_1 - 1}{\varepsilon_1 + 1}\right)$ is necessary to ensure the proper branchcut choice in lossless case, with an infinitesimally small amount of loss/gain the solution will automatically collapse to the correct branch.

The general form of polarizability in transverse polarization, similar to one reported in Eq. (2) for the longitudinal case is:

$$\alpha_t = \frac{\pi^2[\varepsilon_1 + \varepsilon_2 - 6] - 12(\varepsilon_1 + \varepsilon_2)\left(\text{Li}_2(\varepsilon^-) + \text{Li}_2(\varepsilon^+)\right)}{1.5\pi^2(2 + \varepsilon_1 + \varepsilon_2)}$$

$$\varepsilon^\pm = -\frac{(1+\varepsilon_1)(1+\varepsilon_2)(\varepsilon_1+\varepsilon_2)}{\varepsilon_2 \pm \sqrt{-(\varepsilon_1-\varepsilon_2)^2(2+\varepsilon_1+\varepsilon_2)(\varepsilon_1+\varepsilon_2+2\varepsilon_1\varepsilon_2) + \varepsilon_1\left[1+\varepsilon_2(4+\varepsilon_1+\varepsilon_2)\right]}}$$

(S6)

## 3. Dynamic Response

For a silver nanoparticle, we have studied the effect of intrinsic material loss and retardation on the absorption coefficient in the main paper. Here we show further details on these two effects by analyzing the case in which gold instead of silver is considered as plasmonic material and considering separately the effects of radiation and material loss. Conjoined half-cylinders with diameter $2a = 40nm$ are considered, in which the upper half is made of gold with $\varepsilon_\infty = 1.53$, $\omega_p = 2\pi \times 2069\,\text{THz}$, and $\Gamma = 2\pi \times 17.64\,\text{THz}$ based on experimental measurement data. As in the main paper, three configurations are studied separately: $\varepsilon_2 = 1$, $\varepsilon_2 = 3$, and a full gold cylinder for comparison. Fig. S1 shows absorption efficiency (AE) versus frequency for three different scenarios: in the first two panels we consider both realistic losses and retardation effects. Panel (c) shows AE for transverse excitation neglecting retardation but including realistic losses. Panel (d) on the other hand includes the retardation effect but assumes $\Gamma = 0$ (lossless gold).

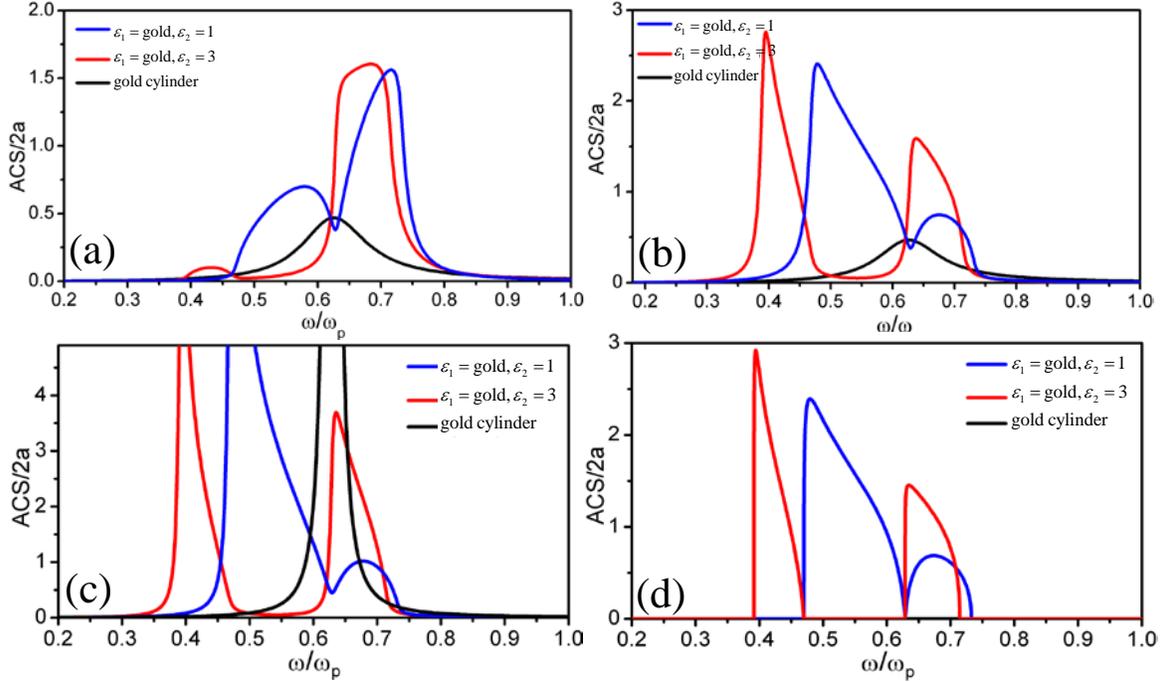

FIG. S1 Absorption efficiency for different configurations under (a) longitudinal excitation: the upper half-cylinder is gold while the lower part either $\varepsilon_2 = 1$ or $\varepsilon_2 = 3$. A full cylinder composed of gold is also included for comparison. (b) Same, but for transverse polarization. (c) Absorption efficiency for transverse excitation under the quasi-static approximation; (d) same as (b), but neglecting gold losses.

Compared to silver (Fig. 4 in the main paper), gold provides slightly lower absorption due to damping of the plasmonic resonance near the corners in presence of a larger material loss. This can be explained also inspecting Fig. S1(d), in which we totally neglect material loss. Following the discussions in the main text, inside the resonance range this particle shows finite absorption in the limit of infinitesimally small losses, slightly dampened by adding realistic material losses. In general, with conventional low-loss plasmonic materials (e.g., silver and gold), the focusing effect still dominates the absorption features of these particles. The effect of retardation can be observed in Fig. S1(c). By including scattering loss, as expected, the absorption efficiency is broadened and dampened. It is interesting that in the case of a single full cylinder, scattering loss affects the total absorption much more drastically than in the composite configurations.

## 4. Field Distributions

Figs. S2a-b show the field distribution in the composite nanoparticle for $\varepsilon_1 = -1.1$ and $\varepsilon_2 = 1$ under longitudinal polarization excitation. These field distributions are calculated analytically as $\mathbf{E} = -\nabla \varphi$ based on the potential distribution obtained in the main paper. Plasmonic oscillations around the corners (Fig. 3a-b main text) result in enhanced fields, which may mathematically become infinite at the edge point in the lossless case for an ideal corner. The potential variation is plotted along the common diameter of the particle for a different example ($\varepsilon_1 = -2, \varepsilon_2 = 1$) in Fig. S2(c).

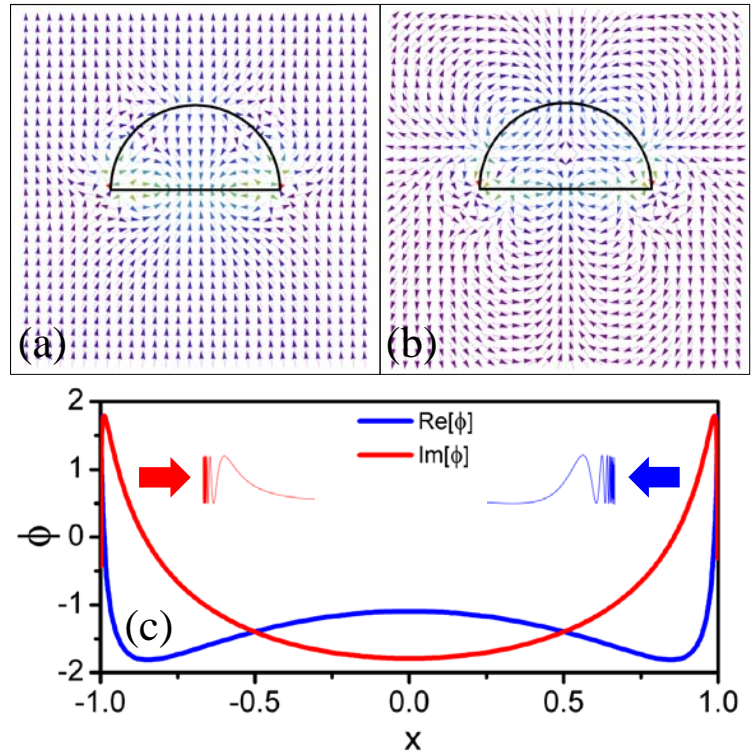

FIG. S2 Electric field distribution for longitudinal excitation: (a) real part in case of $\varepsilon_1 = -1.1$, $\varepsilon_2 = 1$ (b) imaginary part. (c) Potential distribution along the $x$-axis for $\varepsilon_1 = -2$, $\varepsilon_2 = 1$

The different behavior between $\varepsilon_1 = -1.1$ (reported in the main text) and $\varepsilon_1 = -2$ (here) can be interestingly explained considering the wedge solution. It can be shown that for values of $\varepsilon_1$ near $-1$ the frequency of spatial oscillations of eigenmodes at the corner is much larger compared to $\varepsilon_1 = -2$, resulting in oscillations extended farther from the corners. For these situations the field enhancement may be extended more broadly all over the particle, with interesting possibilities to enhance more effectively optical nonlinearities.

## 5. Realistic geometry

In the case of a realistic geometry without singular corners, the absorption efficiency always tends to zero as losses are made very small, as expected. This is also the case if we simulate the ideal geometry of the composite nanoparticle with finite-integration methods (blue curve in Fig. 3d, main text), as a finite curvature is automatically introduced by numerical meshing. However Fig. 3d shows that, despite the rounding, highly oscillatory fields are still induced around the sharp corners, significantly boosting the absorption efficiency even for very low loss materials. The limit for which the resonance effect disappears for low-loss materials depends on the sharpness of the corner we can consider. Our full-wave simulations confirm the robustness of this phenomenon on the corner curvature and edge bluntness, consistent with previous results for other types of plasmonic resonances [2]. For sharper edges, the number of quantized resonances increases and the overall effect gets closer the ideal solution [3]. Figs. S3a-b show the field distribution for a silver half-cylinder having an ideal corner using our analytical solution in the quasi-static limit and for a blunted structure with $2a = 40 nm$ and $r = 2 nm$ using full-wave simulations at $f = 925\,\text{THz}$ for longitudinal excitation. Fig. S3(c) also shows the power flow in the blunted case.

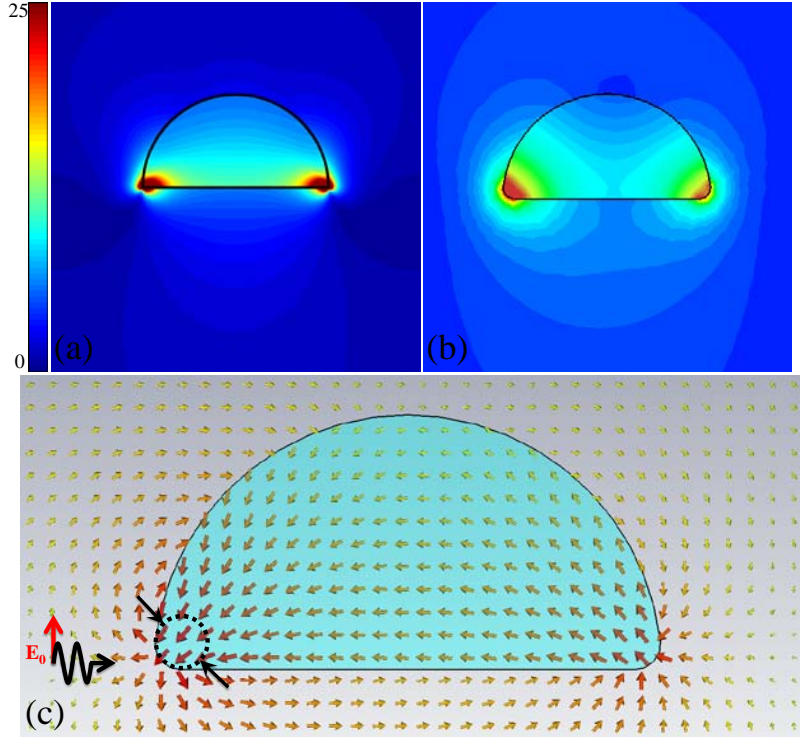

FIG. S3 Field distribution on (a) an ideal silver half cylinder with $\varepsilon_1 = -0.5287 - 0.026j$ and (b) blunted configuration at the corresponding frequency. (c) Power flow

Interestingly, even with a relatively large edge curvature (and including scattering losses), field enhancement is pretty comparable with the ideal case. The small asymmetry in the distribution is due to the direction of the impinging wave, but since the particle is small compared to wavelength, this effect is almost negligible. Power flow is plotted in a log100 scale, implying large power concentration inside the particle, responsible for large absorption efficiencies.